# A Parametric Study on the Starting Transients of a Vacuum Ejector-Diffuser System

Ankit Mittal[1] and Rajesh G[2†], H. D. Kim[3] and V[4]. Lijo

[1]Scientist, VSSC, *Indian Space Research Organization*

[2†]Corresponding Author, Assistant Professor, Keimyung University, South Korea

[3]Professor, Andong National University, South Korea

[4]Associate Professor, College of Engineering, Trivandrum, Kerala, India

**In this endeavor the transients persisting in a vacuum ejector is studied by numerically simulating the flow field, and experimentally validating the simulated results. An inertial effect was discovered in the study due to which the direction of mass flux changes (hence the recirculation zone moves forward and backward) during the transients and the pressure in the secondary chamber rises and falls, this behaves as a damped oscillatory flow in which the direction of mass flux keeps on changing but it finally settles to position where there is no mass flux in either direction. The movement of recirculation zone is studied here. The affect of the thickness of the primary and secondary jet and the affect of increase in secondary chamber volume on the inertial effect is studied. The results obtained also shows that the pressure and mass flux through the secondary chamber depends highly on both the parameters. The inertial effect reduces with the reduction in thickness of primary and secondary flow. As the volume of the chamber increase the inertial effect decrease further. Later the affect of primary pressure is studied on the inertial effect, the inertial effect decreases as the primary jet pressure increases, given that all other parameters are kept constant. Experiments are also done to validate the present study.**

## NOMENCLATURE

R   = thickness of primary and secondary jet
$M_p$ =mass through primary nozzle
$M_s$ =mass flux through secondary nozzle
M   =Mach number
$V_i$ =volume of secondary chamber
P   =pressure
T   =temperature
E   =internal energy
$u_i$ =velocity componenets

# I. INTRODUCTION

THE vacuum ejector system is a simple device which is used to transport low pressure fluid by using a high pressure jet. It consists of a primary chamber (and nozzle), secondary chamber, mixing section and the diffuser [1].

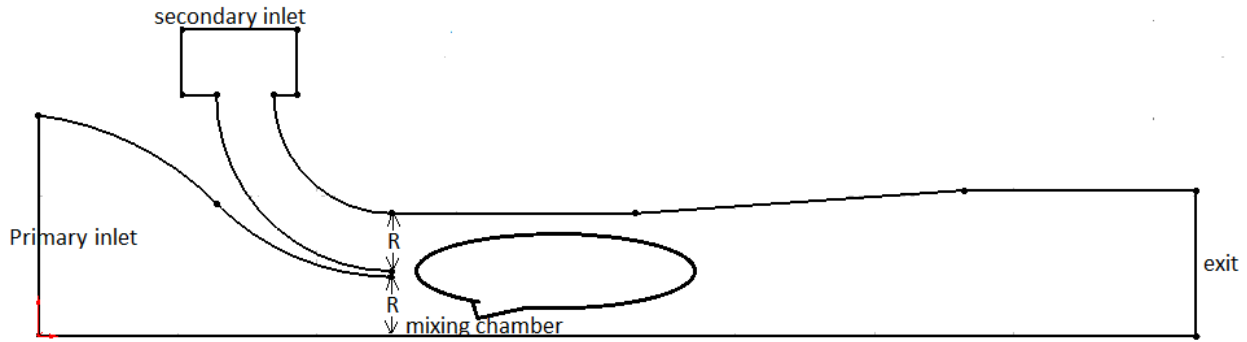

**Figure 1: Vacuum ejector system**

The working is based on a shearing action caused by the flow coming out of primary nozzle on the secondary flow, that is, the primary flow drags out the secondary flow emptying the secondary tank. It has a lower efficiency than other fluid transport devices [4], but due to its simple design with no moving parts it proves itself useful, and it needs little maintenance.

Ejector system can be used for Thrust augmentation, the secondary flow augments the direction of flow and hence the direction of thrust (V/STOL system) [5, 6]. It can also be used in high altitude test facility [7], a high altitude facility is installed to conduct tests at lower pressure and density, and hence vacuum ejector system can be used to draw air from secondary chamber and creating required partial pressure. Can also be used in combustion facility [8], the refrigeration system [9], natural gas generation [10], fuel cells [11], noise-control facility [12] etc.

In much of the earlier works [3, 9 10, 13], except [2], the secondary chamber was assumed to be of infinite size that can supply mass indefinitely, hence steady flow assumption was very well valid. However in all practical applications the secondary chamber has a finite volume, hence the steady state assumption breaks down. In the practical scenario the flow field developed is really important from the view of designing the ejector system. Initially the secondary mass entrains due to the shearing action of primary flow and the pressure in the secondary chamber reduces. However after sometime the shearing action continues but the pressure in the secondary tank hardly reduces. A recirculation zone appears in the mixing chamber which allows for constant secondary chamber pressure while the shearing action continues. But some questions remain unanswered, like, how does the recirculation zone behave in the ejector? Why does it move? How the steady state result and transients change as volume of secondary chamber increase? Does the jet thickness (R) affect the flow field? How the primary jet pressure influence the transients? All these questions are addressed in this study.

## II. METHODOLOGY

Primary flow transient is very small as compared to secondary flow transient; hence it is neglected in this study. The numerical solution is to be carried out for different volumes of secondary chamber, and the transient flow characteristics are studied, some experiments are to be conducted to validate the simulation results. For the numerical simulations steady state simulations will be computed initially with secondary chamber walls open, when the desired convergence is reached the unsteady simulation will be computed with chamber walls closed. To save computational time the secondary chamber is considered to be very small and connecting tube itself is assumed to be the secondary chamber.

### 1. COMPUTATIONAL METHODOLOGY

Systematic grid independence studies have been performed in this study for qualifying the grid size. Three separate grids were generated with 27000 cells, 47000 cells and 97000 cells. The grids are shown in figure 2 and 3. Figure 4 demonstrates the velocity magnitude along the axis for three different grids. The grids have been made so that the near wall thickness has a y+ value near 1 which is required by the turbulence model for resolving the flow field.

The time step study for the transient flow has also been study systematically with three different time steps 1e-7 sec, 1e-6 sec, 1e-5 sec. Figure 5 shows the mass flux through the secondary chamber and pressure in the secondary chamber with time for different time steps. The conclusion from grid independence and time step independence study is that the grid with 47000 cells and the time step of 1e-7 seconds is most optimized for the given problem.

Ejector geometry is shown in figure 1 with the boundary conditions. The axis symmetric, coupled implicit solver is chosen for the steady as well as unsteady simulations with boundary condition at inlet being the Pressure inlet condition with the total pressure of 3 bars and total temperature of 300 K. At secondary inlet also pressure inlet boundary condition is given with total temperature as 1 bar and total temperature of 300 K. The boundary condition at exit is pressure outflow with back pressure imposition of 1 bar; all the walls are adiabatic with no slip condition. Dry air is used as working fluid and viscosity is calculated using Sutherland equation.

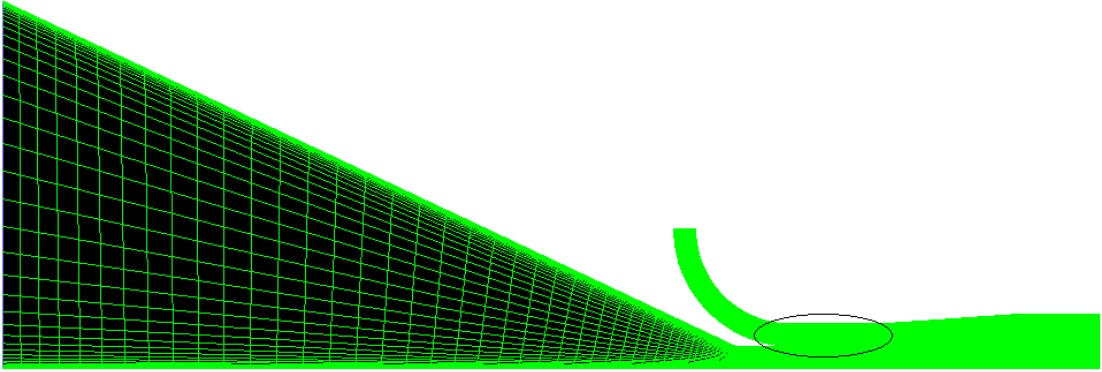

Figure 2: Computational grid

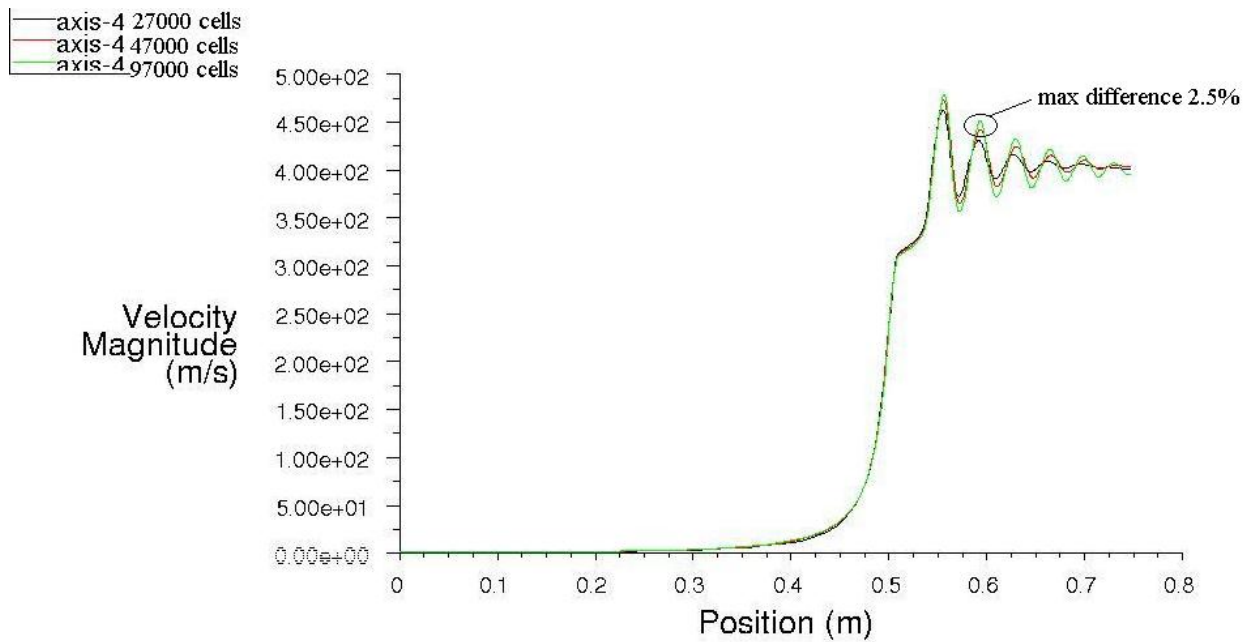

Figure 3: Velocity magnitude for three different grids.

Due to inherent turbulent nature of problem the turbulence model chosen plays a very important role. The ideal turbulent model would be the LES but prohibitively large number of grid cells increases the computational time immensely, hence limiting the use of LES model. The eddy viscosity model does not predict the wall shear stress very well because the Reynolds stresses are modeled in this approach. The Reynolds stress model proves to be the best model for the problem in hand as it takes less computational time as well as it predicts the wall shear stress better as the actual Reynolds stress transport equation is numerically solved here.

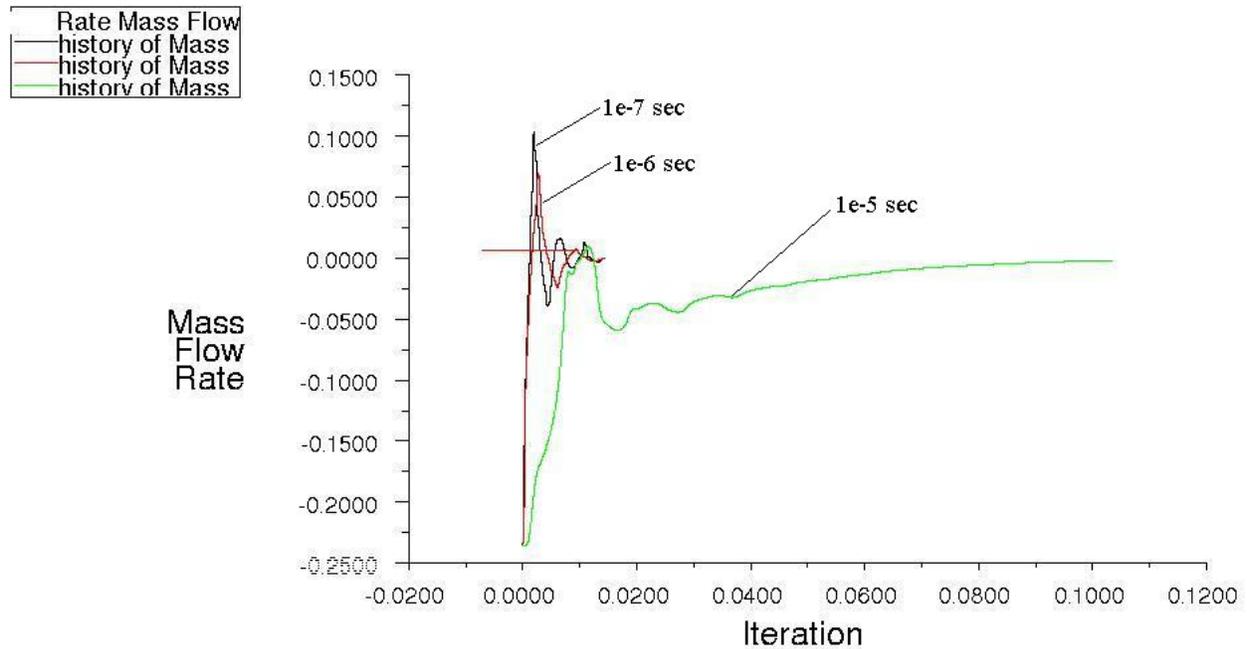

**Figure 4: Mass flux history in secondary chamber with different time steps.**

With these conditions the steady numerical calculation is started with an assumption that infinite mass supply from secondary chamber is available. The simulations are stopped when the net mass flux becomes approximately of the order of 0.2% of the inlet mass flux, and the solution is no more changing.

For the numerical simulation of the experimental setup the grid was created with the cell thickness near the wall is 2.4 μm, and total number of cells are >400,000 cells. Compressible viscous N-S equation was used with a RSM turbulence model.

The same boundary conditions are given as the axis symmetric model, except that the lower surface which is wall. The steady simulation is run until the net mass flux reaches 1e-4 Kg/s, and the residue drops to 1e-5 level. When the above described level of convergence is obtained the unsteady simulation is started with the secondary inlet wall closed.

## 2. GOVERNING EQUATIONS

The flow field in ejector is governed by compressible axis symmetric Navier Stokes equation. The equations below however are written in Cartesian co-ordinates for easiness. In this study Favre averaged NS equations are used because of their pertinence in density varying flows. The equations hence are written below.

### Continuity

$$\frac{\partial \rho}{\partial t} + \frac{\partial}{\partial x_i}(\rho u_i) = 0 \qquad (1)$$

### Momentum

$$\rho \frac{D(u_i)}{Dt} = -\frac{\partial P}{\partial x_i} + \frac{\partial}{\partial x_j}\left[\mu_{\text{eff}}\left(\frac{\partial u_i}{\partial x_j} + \frac{\partial u_j}{\partial x_i} - \frac{2}{3}\delta_{ij}\frac{\partial u_k}{\partial x_k}\right)\right]$$
$$+ \frac{\partial}{\partial x_j}\left(-\rho \overline{u'_i u'_j}\right) \qquad (2)$$

The velocity here is mean velocity (mass averaged), while the prime components are the disturbance components (due to turbulence) and it denotes reynolds stresses. E and T are also mass averaged, while $\tau_{ij}$ is the stress tensor

$$\frac{\partial}{\partial t}(\rho E) + \frac{\partial}{\partial x_i}[u_i(\rho E + P)]$$
$$= \frac{\partial}{\partial x_i}\left[\left(\alpha + \frac{C_p \mu_t}{P_{rt}}\right)\frac{\partial T}{\partial x_i} + u_j(\tau_{ij})_{\text{eff}}\right] \qquad (3)$$

Where,

$$\tau_{ij} = \mu_{\text{eff}}\left(\frac{\partial u_j}{\partial x_i} + \frac{\partial u_i}{\partial x_j}\right) - \frac{2}{3}\mu_{\text{eff}}\frac{\partial u_i}{\partial x_i}\delta_{ij} \qquad (4)$$

The ideal gas equation is also used for balancing the equations with unknown

$$\frac{P}{\rho} = rT \tag{5}$$

The transport equation for Reynolds stress, which are computed for Reynolds stresses in RSM model is

$$\frac{\partial}{\partial t}\left(\rho \overline{u'_i u'_j}\right) + \frac{\partial}{\partial x_k}\left(u_k \rho \overline{u'_i u'_j}\right) = D^T_{ij} + D^L_{ij} + P_{ij} + \phi_{ij} + \varepsilon_{ij} \tag{6}$$

The right hand side denotes turbulent diffusion, molecular diffusion, stress production, pressure strain, and dissipation respectively, where,

$$D^L_{ij} = \frac{\partial}{\partial x_k}\left(\mu \frac{\partial}{\partial x_k} \overline{u_i u_j}\right) \tag{7}$$

$$P_{ij} = -\rho\left(\overline{u_i u_k}\frac{\partial u_j}{\partial x_k} + \overline{u_j u_k}\frac{\partial u_i}{\partial x_k}\right) \tag{8}$$

The equations for other parameters are not known and have to be modeled, the modeled equations are,

$$D^T_{ij} = -\frac{\partial}{\partial x_k}\left[\rho \overline{u_i u_j u_k} + \overline{P\left(\delta_{kj} u_i + \delta_{ik} u_j\right)}\right] \tag{9}$$

$$\phi_{ij} = \overline{P\left(\frac{\partial u_i}{\partial x_j} + \frac{\partial u_j}{\partial x_i}\right)} \tag{10}$$

$$\varepsilon_{ij} = -2\mu \overline{\frac{\partial u_i}{\partial x_k}\frac{\partial u_j}{\partial x_k}} \tag{11}$$

## 3. EXPERIMENTAL SETUP

The experiments were carried out to validate the numerically simulated results on two different experimental setups to reduce the experimental inaccuracies. Schlieren pictures were obtained of the flow field in the mixing chamber.

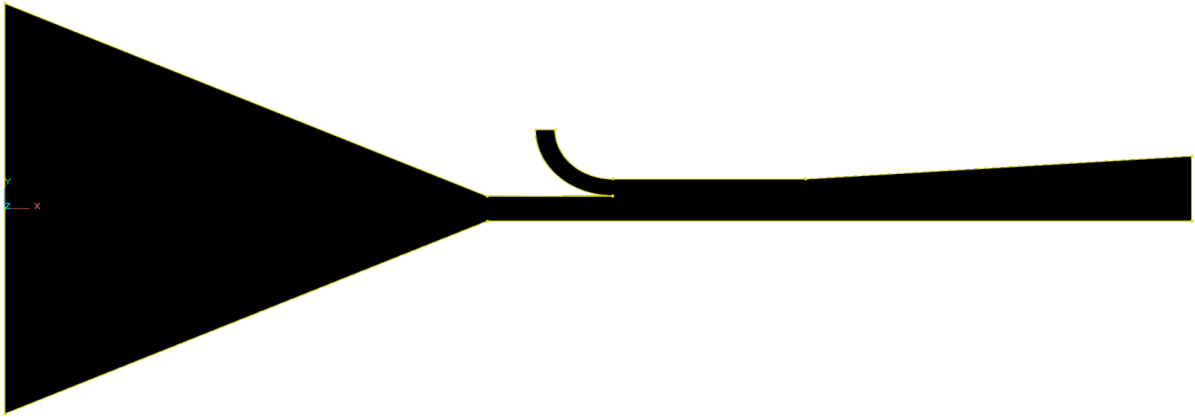

**Figure 5: cross sectional view of experimental setup.**

Schlieren is the most advanced and state of the art technology to visualize the flow. Unlike many other methods of flow visualization in which physical material is put in the experimental chamber, in Schlieren light properties are used to visualize the flow. The setup has a light source, two highly polished mirrors, a screen and a knife edge.

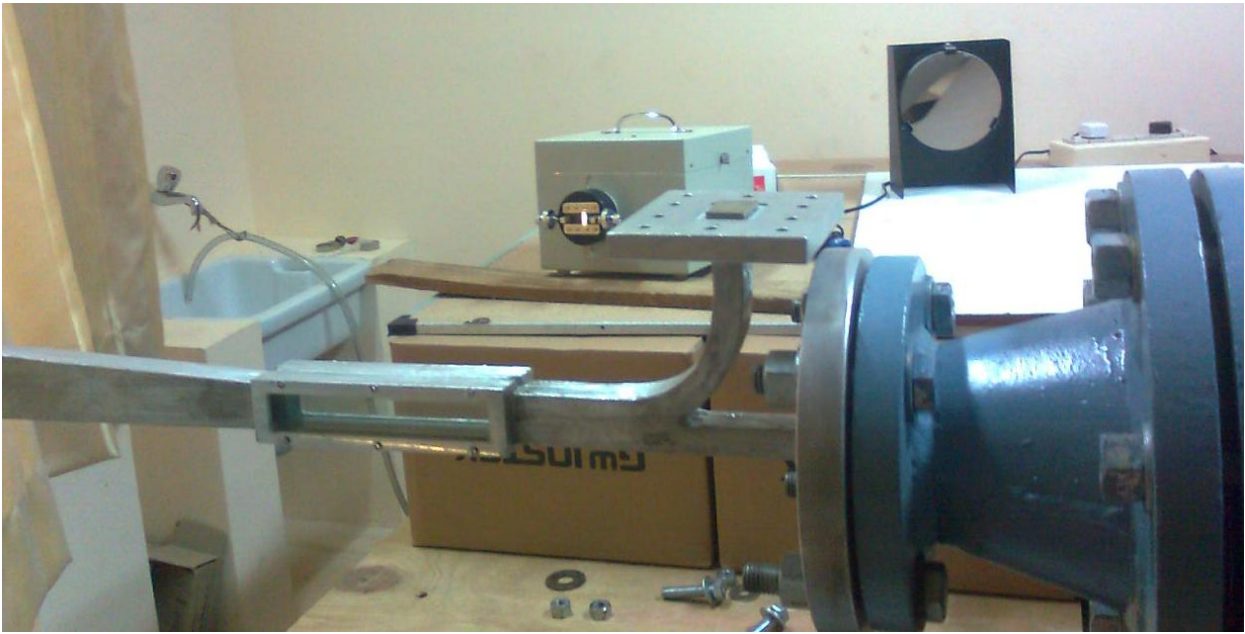

**Figure 6: 2nd model with Schlieren setup**

The steady state pressure measurement was also done in the secondary chamber, and the pressure at the exit was also measured to check the pressure imposition condition of 1 bar. The simulations were carried out at 1 bar atmospheric pressure, while the atmospheric pressure here is less than 1 bar, hence this is source of some errors.

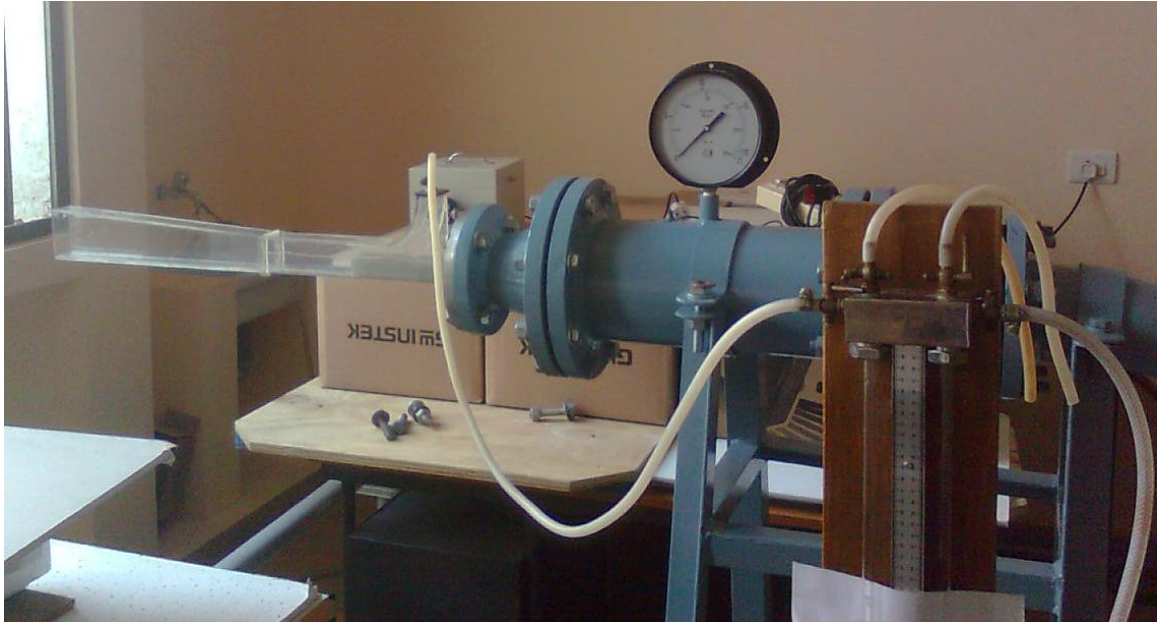

**Figure 7: The experimental setup at the facility with manometers attached.**

### III. RESULTS AND DISCUSSION

**1. Part 1- Axis symmetric Vacuum ejector body results**

Figure 8(A) shows the pressure variation is the secondary and mixing chamber with the time (for a particular volume and R) during the transient starting process of the ejector system. Point 'a' marks the beginning of the transient and point 'f' marks the ending of the transient phase and starting of the steady flow. Figure 8 (B) shows the mass flux history through the secondary chamber for the same volume and R, it can be seen that the time instant (13.8 ms) when mass flux reaches a constant value is same as the time when pressure reaches the constant value.

Figure 10(A), (B) shows the mach contours and the velocity vectors, it can be seen that the air in the secondary chamber is dragged out by the shearing action caused by the primary jet (figure 10 (B)), as the ejector reaches a steady condition the pressure in the secondary and hence the mixing chamber falls and the primary jet keeps on expanding as can be seen in Figure 12. It should also be stated that at the steady state the pressure in secondary and mixing chamber becomes same but due to the recirculation zone the fluid is still dynamic and mass entrainment continues.

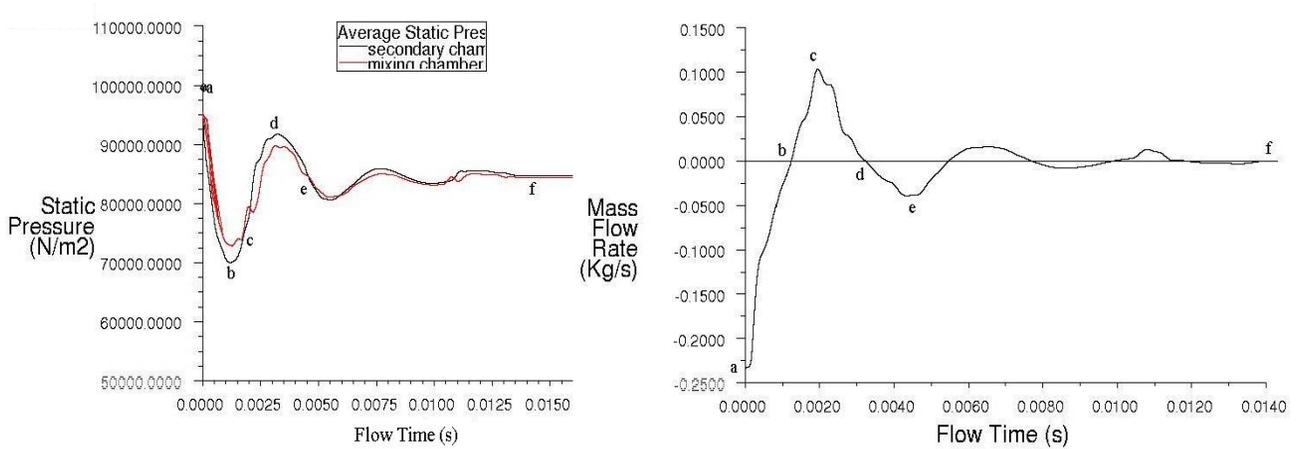

**Figure 8: A) Pressure in secondary chamber and mixing chamber. B) Mass flow rate through secondary chamber.**

After about 0.63 ms a recirculation zone appears (formation of recirculation zone is reported in [2] also) at the ejector exit (Fig 13(A)) and a small recirculation also appears in the diffuser, as the time increases the recirculation expands inwards to the mixing chamber, after further more time the recirculation zone expands till the secondary chamber (Fig 13(D)), at about 1.35 ms the recirculation moves in the secondary chamber (Fig 13(E)). An interesting phenomenon is seen as the time increases further, the recirculation drifts backward to the exit of the secondary chamber and then to mixing chamber (Fig 13(F)). Again as the time increases the recirculation zone moves in the secondary chamber, and after oscillating few times it finally rests stretching till the secondary chamber as shown in figure 13(G). This movement of the recirculation zone back and forth is due to the inertial effects and can be understood very well from the pressure and mass flux rate through the secondary chamber Fig 8. Initially the pressure in the secondary chamber is higher than the pressure in mixing chamber hence the mass flows in the mixing chamber from the secondary chamber, as the time increases the pressure in the secondary chamber decreases at a higher rate than the rate of pressure reduction in mixing chamber and hence the rate of mass flow reduces. The pressure in mixing chamber surpasses the secondary chamber pressure, but still the mass entrainment continues due to momentum of flow, finally when the pressure difference overcomes the momentum of flow (point b) the mass flow rate reverses (the mass start to move in the secondary chamber) and hence the pressure in secondary chamber starts to increase. Now after the pressure in secondary chamber goes much higher than the pressure in the mixing chamber (to overcome the momentum of fluid) the mass flux direction again reverts back, this phenomenon continues till the flow reaches steady state , and this inertial effect is responsible for oscillation of recirculation zone. The Mass flux through secondary chamber and the pressures in mixing and secondary chamber for different R and volumes were also compared and the trend was found to be similar. The formation of recirculation zone is the only condition which would allow for constant pressure in secondary chamber though the mass entrainment continues. The steady state assumption is valid only after the recirculation zone appears and it stops *oscillating*.

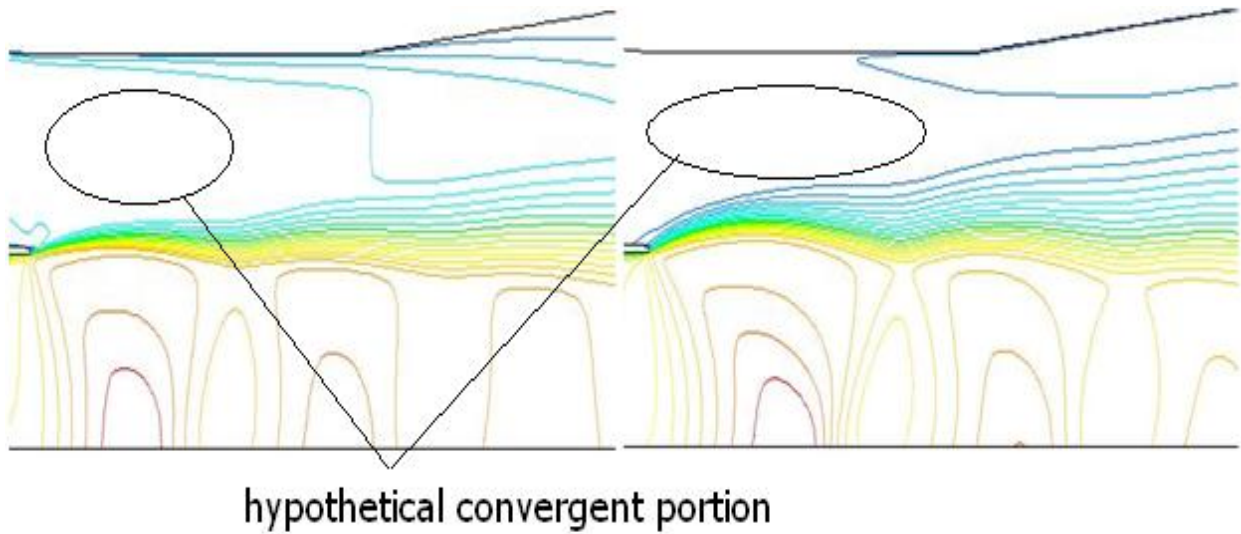

**Figure 9: velocity contour at 0.03ms and 1.13ms, showing expanding primary jet and hypothetical convergent portion**

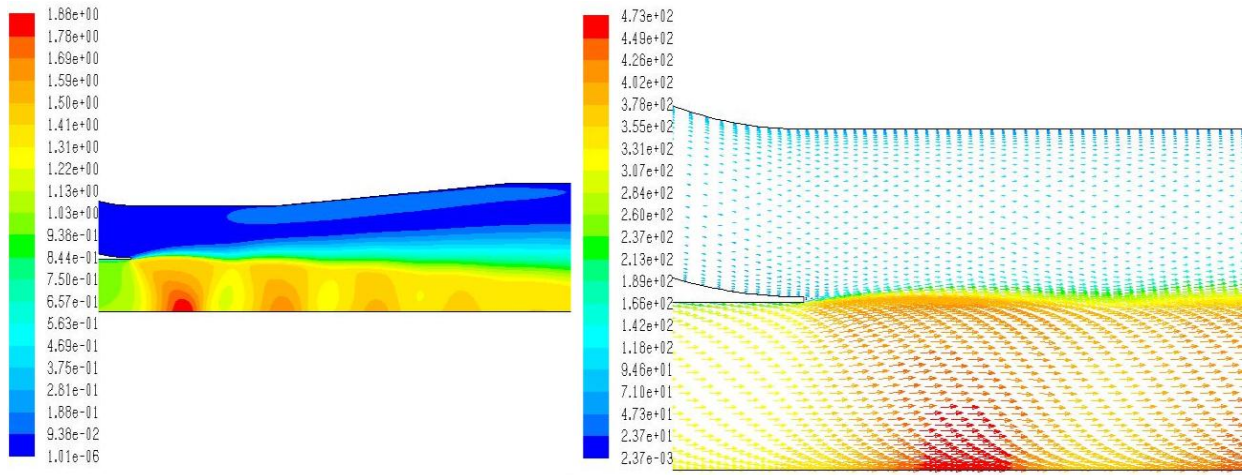

**Figure 10: (A) Mach contour, (B) velocity vectors at 0.03 ms.**

Figure 14 shows the shear stress on the upper wall at different time intervals, after about 0.6 ms the shear stress at the wall has negative sign on some part of wall indicating a recirculation, as the time progresses more part of upper wall has negative shear stress indicating the upstream stretching of the recirculation zone towards the secondary chamber.

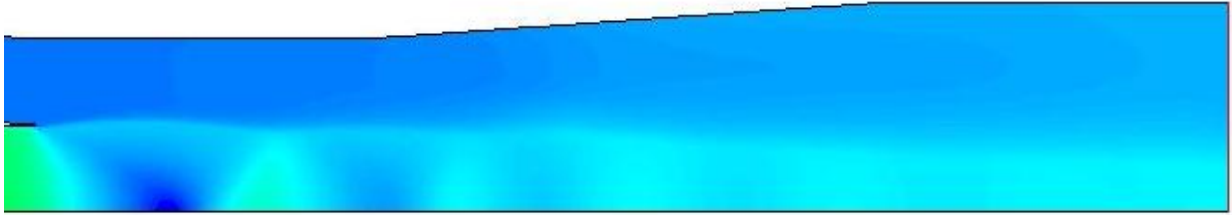

**Figure 11: Density contours at 0.63 ms.**

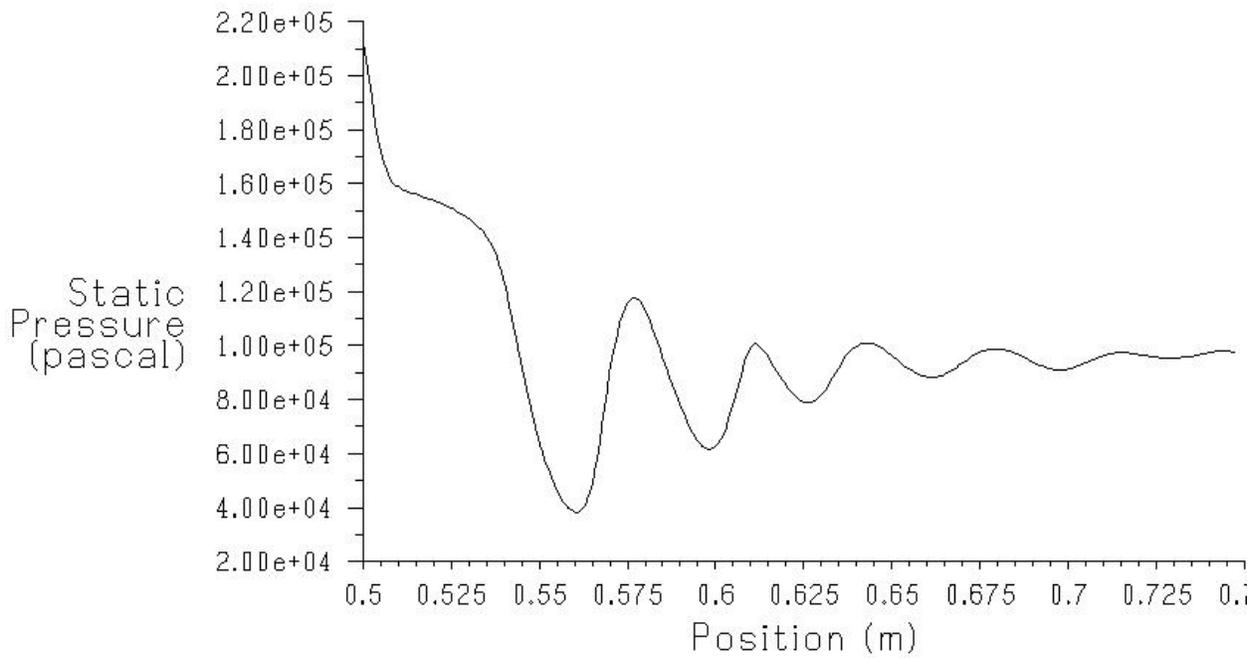

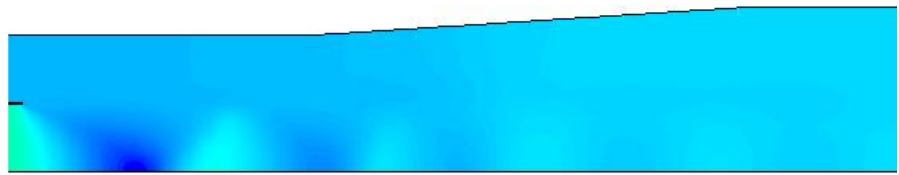

**Figure 12: pressure contours showing pressure rise across mach disk at 0.63 ms.**

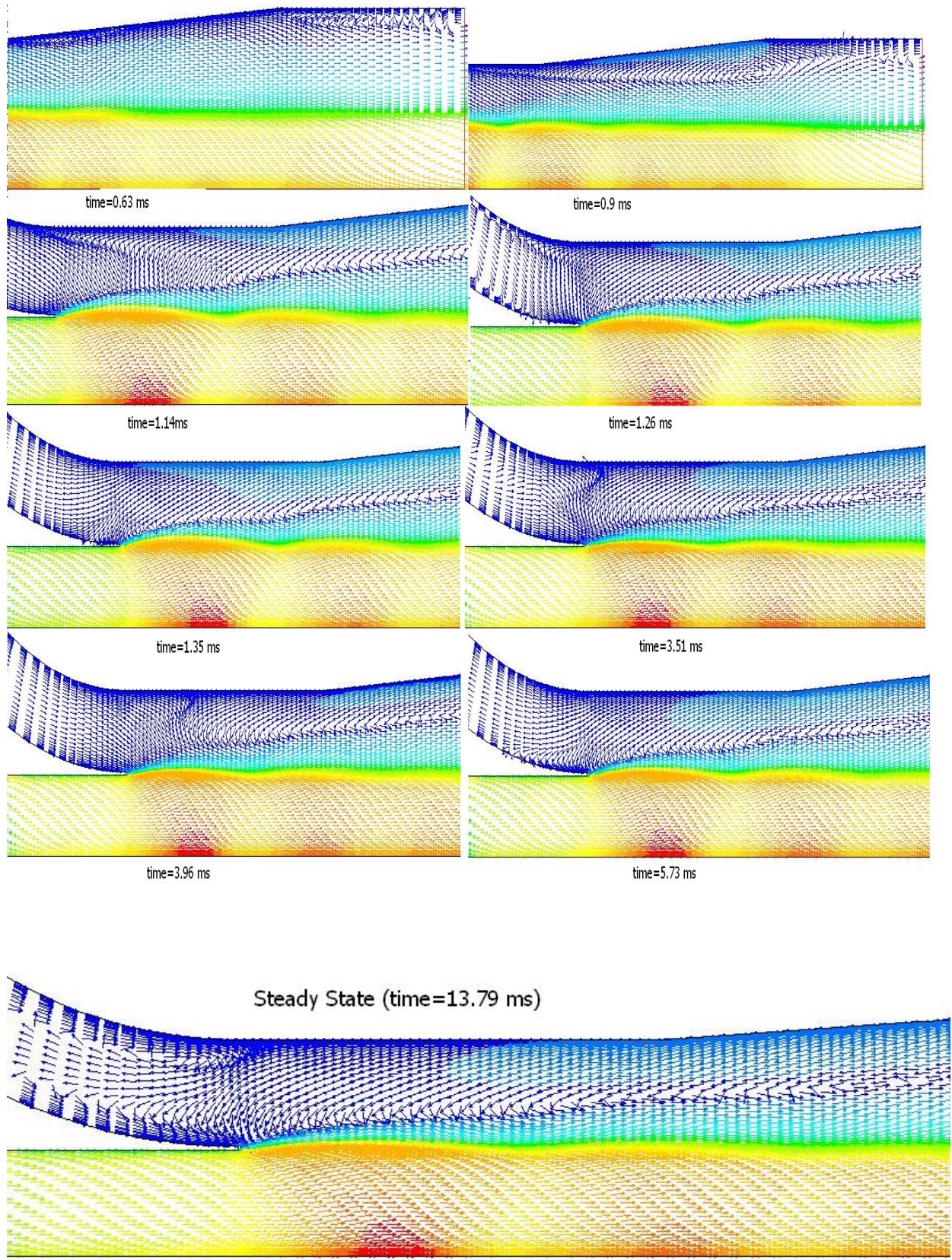

**Figure 13: Position of recirculation zone with time.**

The reason for formation of recirculation zone can be attributed to the pressure rise across a mach disc (Fig 12), that is, the primary flow expands forming a hypothetical converging section for secondary jet as shown in Fig 9, since the secondary jet is subsonic its pressure falls as it passes through the convergent section, but downstream the pressure across mach disk increases, the flow has to work against the adverse pressure gradient, but the momentum of flow is not enough and hence it separates, resulting in a recirculation zone.

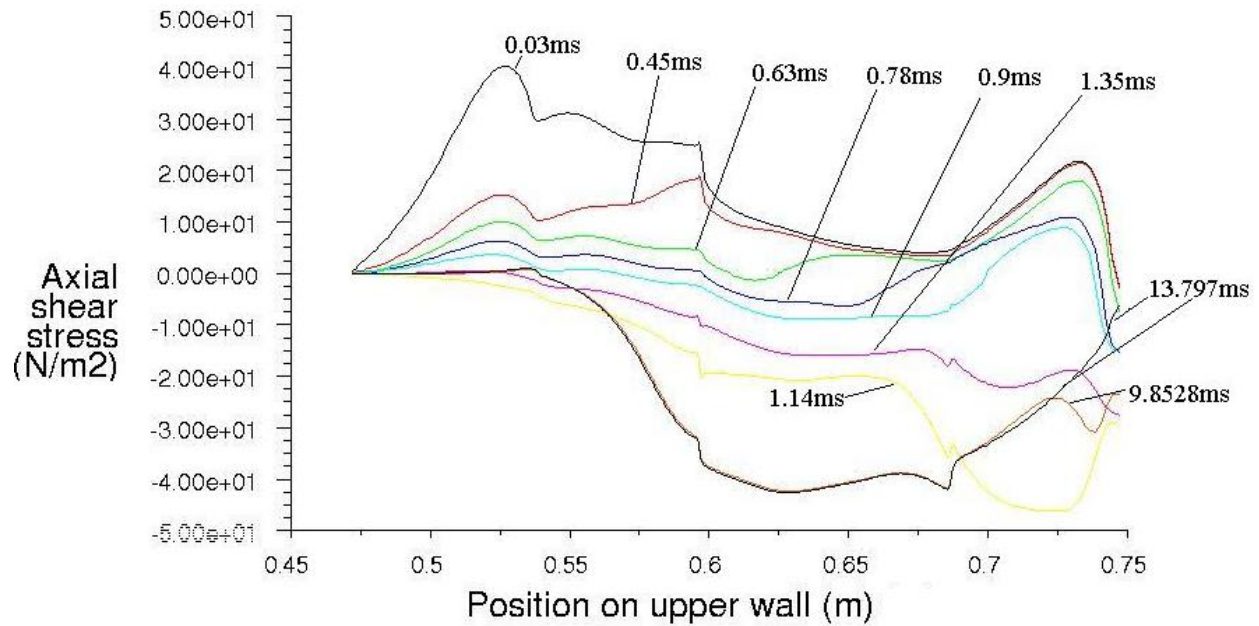

**Figure 14: Shear stress at different time steps.**

The flow was also simulated for different volumes and different values of primary and secondary jet thickness (R). Figure 15 shows the pressure and mass flow rate through the secondary chamber for same R but different volumes (V3>V2>V1), and the volume effect on inertial effect was noted, as the volume increases the inertial effect reduces as can be seen from the plot. This is intuitive as since the volume is large the suction created is less and the pressure in the secondary chamber never crosses the pressure in mixing chamber hence there is no inertial effect in question (figure 16). Also the lowest pressure created increases with the increase in volume.

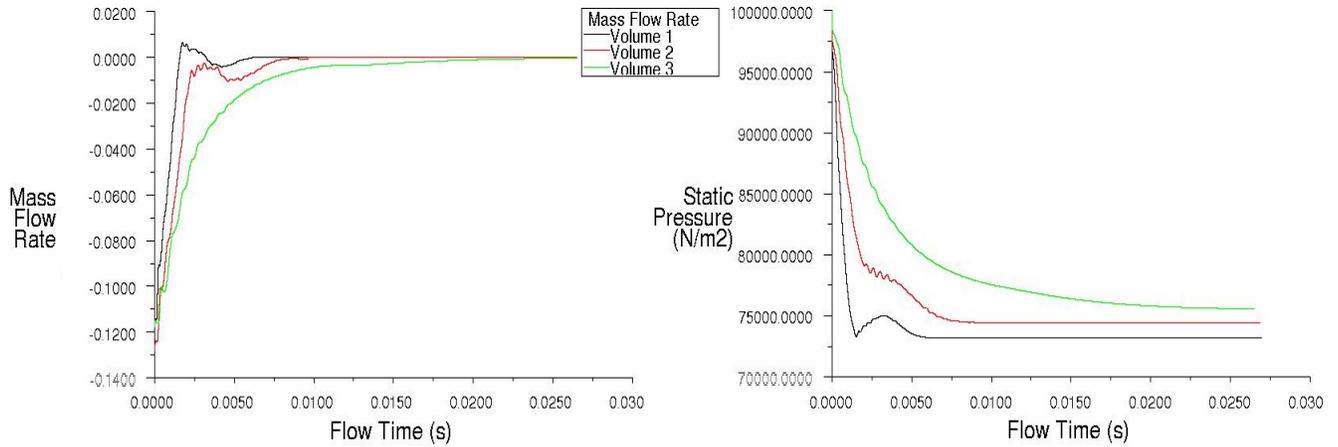

**Figure 15: Mass flux and Pressure through secondary chamber for different volumes and same R**

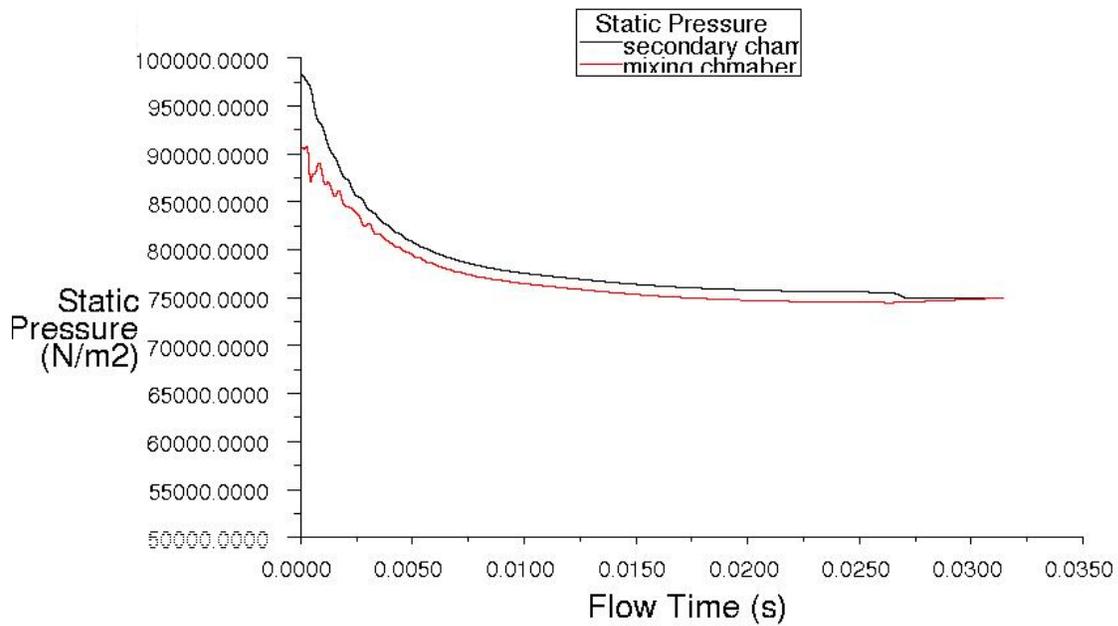

**Figure 16: Pressure in mixing and secondary chamber for Volume V3**

Figure 17 shows the mass flow rates and pressure through secondary chamber for different R, as can be clearly seen from the plots the inertial effect reported before increases as the value of R increases, this happens due to higher momentum rate available with the fluid as R increases. Figure 18 (A, B) shows the velocity vectors at three different sections in the mixing layer and the diffuser at the starting of transients and at 9.3 ms. The figure displays the growth of shear layer and the recirculation zone formed.

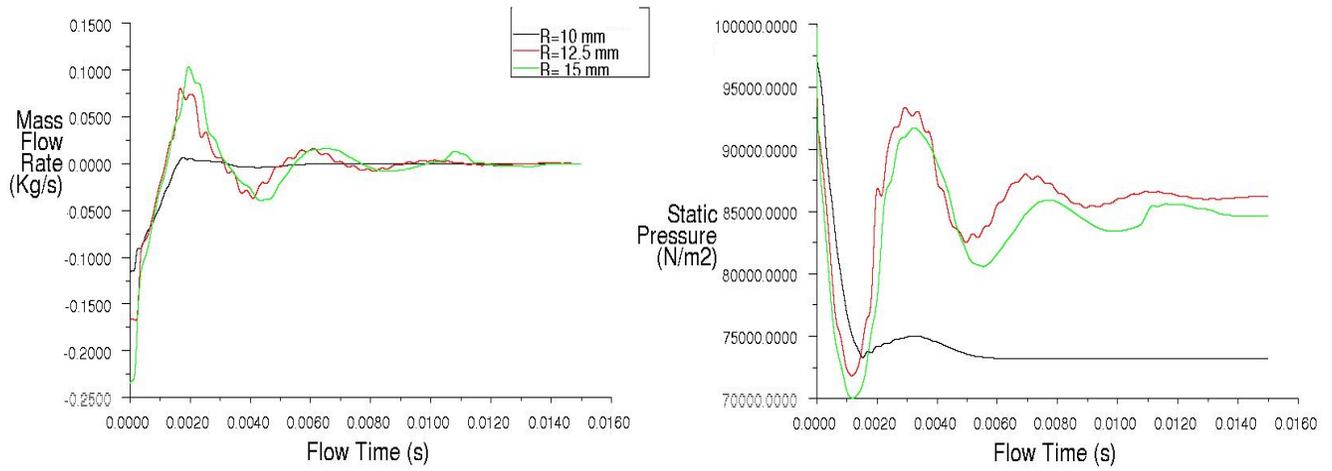

**Figure 17: Mass flux and Pressure through secondary chamber for different R**

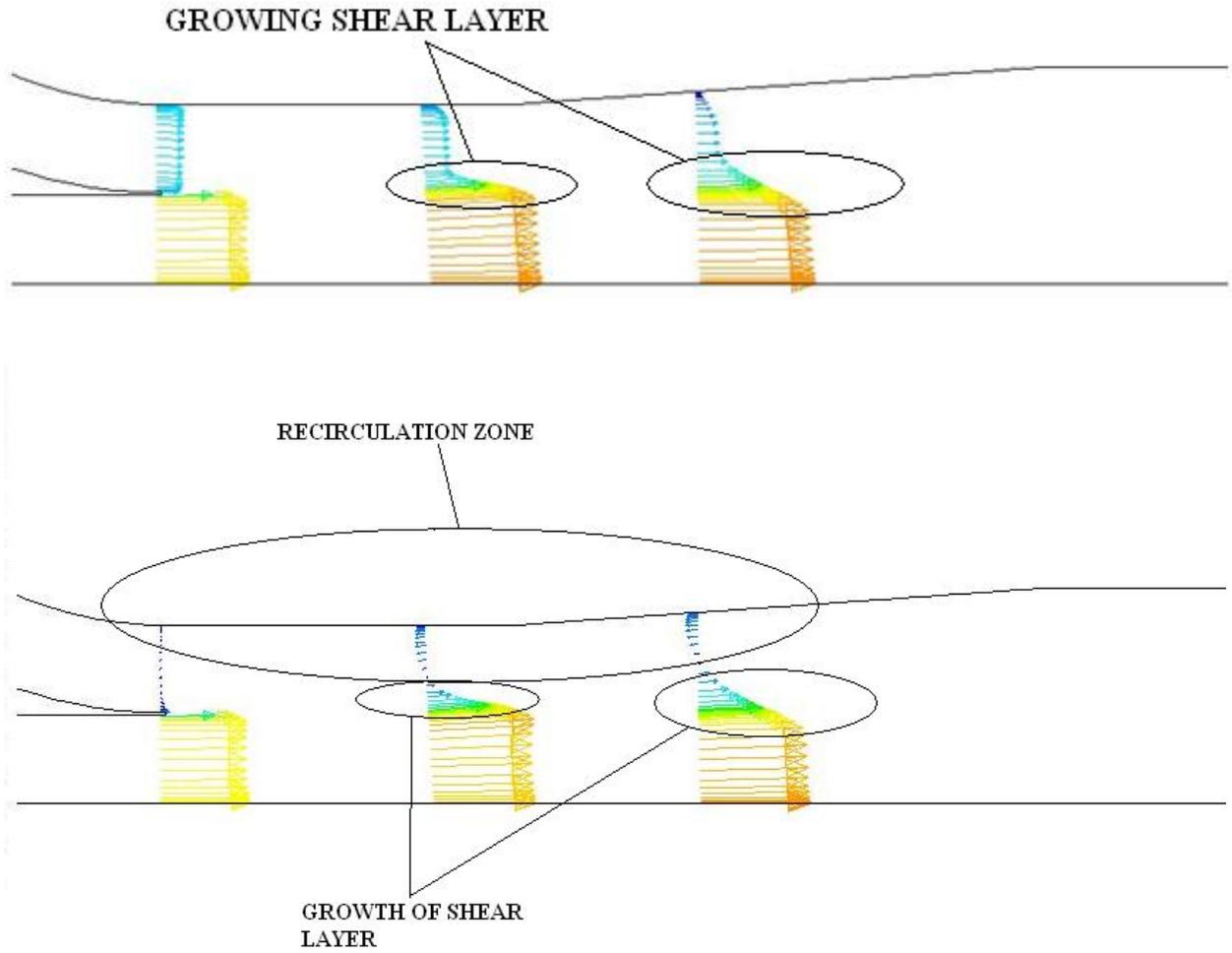

**Figure 18: (A) velocity vectors at 0.03 ms, (B) velocity vectors at 9.3 ms.**

The inertial effect, is also altered by the primary jet pressure, as the primary jet pressure increases the inertial effect decreases and the time taken to reach the steady state also decreases. The inertial effect was compared for three primary jet pressures viz. 3, 3.5, 4 bar, for a volume same volume of secondary chamber (V4), the result can be seen in figure 19. Table 1 shows the steady state attaining time with primary jet pressures, the trend is nearly linear and decreasing.

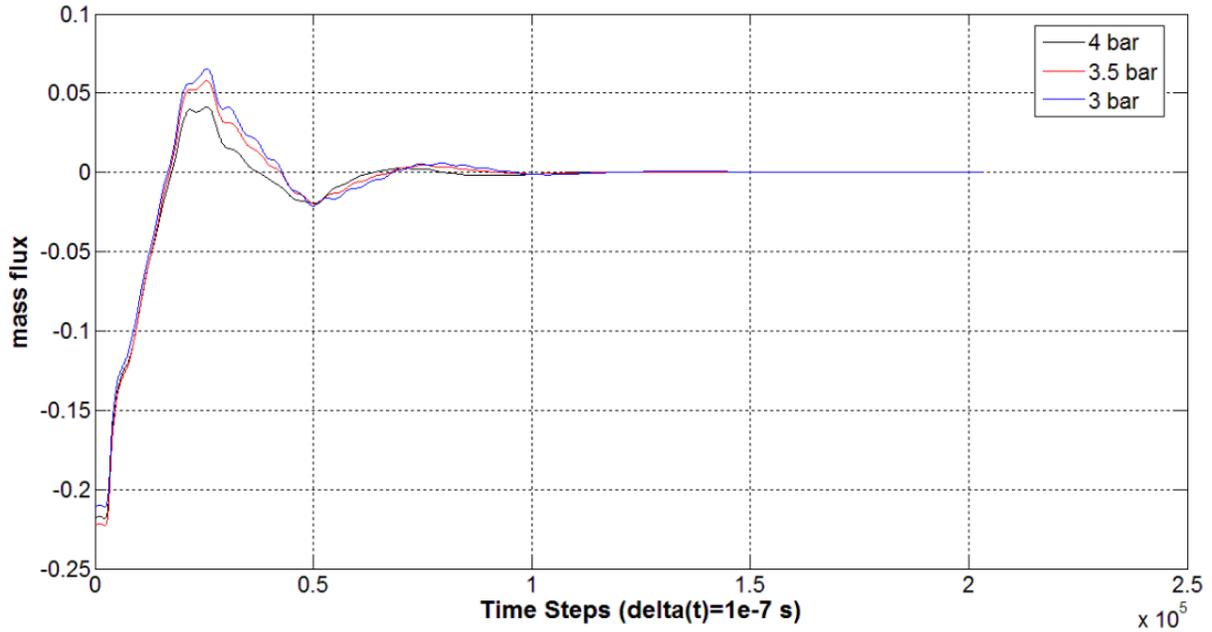

Figure 19: Effect of primary jet pressure on transients.

Table 1. steady state time versus primary jet pressure

| Sl. No | Primary jet pressure | Time taken to reach steady state |
|---|---|---|
| 1 | 3 | 1.8 ms |
| 2 | 3.5 | 1.52 ms |
| 3 | 4 | 1.25 ms |

## 2. Part 2- Experimental Validation

As explained earlier experimental validation was done by

1. Comparing schlieren photo of the flow and density contours of the numerically simulated flow.

2. Comparing the static pressure in the chamber for the numerically simulated flow and experimental result.

The density contour of the numerically simulated flow is superimposed to the schlieren photograph, as can be seen in the Figure 20 and 21 the flow field matches very well. The primary jet which is underexpanded expands when enters in the mixing chamber and its mach number increases. A mach disk is formed due to which the pressure increases and the flow on the lower wall separates. Another oblique shock emanates from the triple point which reflects from the wall.

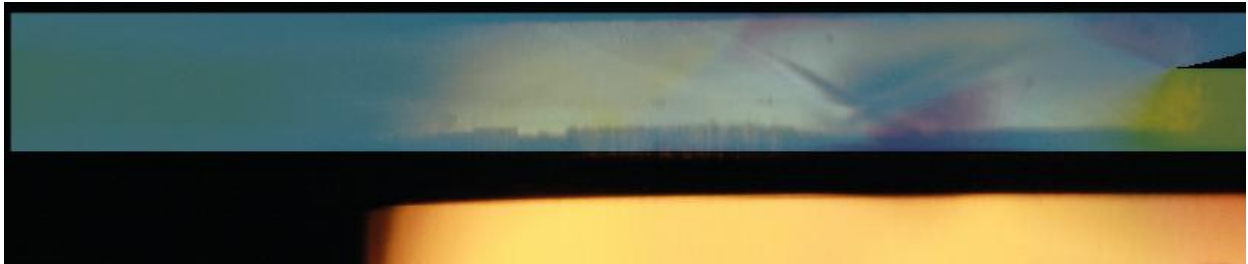

Figure 20: Superimposed flow fields (numerical and experimental)

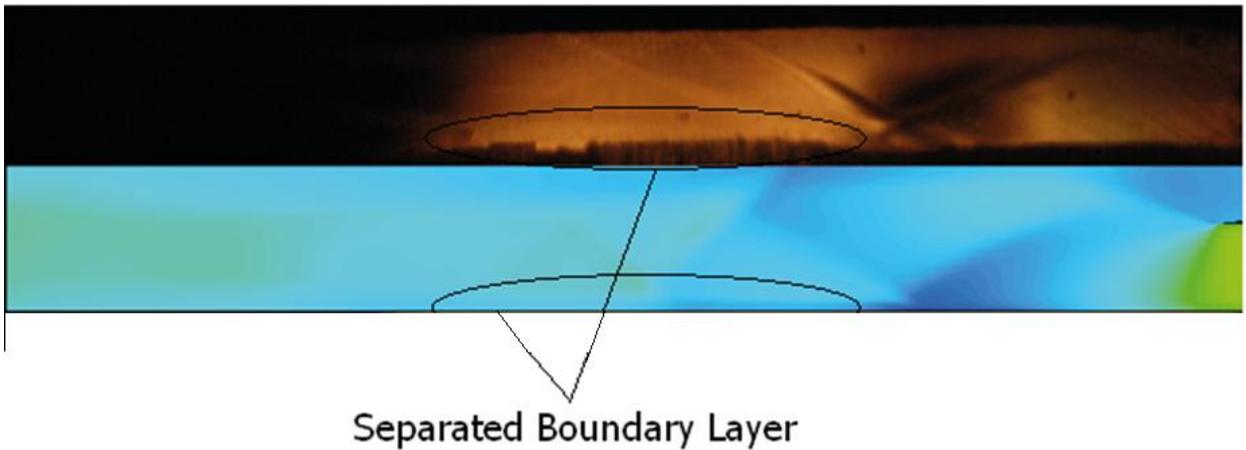

Figure 21: The comparison of density gradients in the flow fields (numerical and experimental)

Table2: Comparison of numerical and experimental data

| Experimental | Numerical |         |
|--------------|-----------|---------|
|              |           |         |
| 0.33 bar     | 0.37 bar  | Model 2 |

The pressures are very well matched. As the primary jet valve is opened some of the fluid in the secondary chamber is already drained, lowering the pressure, causing some difference in the values for experimental pressure and the computational pressure.

# IV. CONCLUSION

The study of transients in a vacuum ejector diffuser system reveals the way in which transient flow achieves the steady state. The steady flow with fixed mass supply secondary chamber is very different from the steady flow with infinite mass supplying secondary chamber assumption. A recirculation forms, and the inertial effects responsible for oscillatory motion of the recirculation zone have been discovered in this study. The steady state assumption of the vacuum ejector system is valid only after the recirculation stops oscillating. The jet thickness (R) governs the inertial affects, as R increases the inertial effect increases due to higher momentum rate of fluid. The volume of secondary chamber also governs the inertial effect, as the volume increases the inertial effect reduces, this is intuitive, since the volume is large, the suction created in secondary chamber is less and the pressure in the secondary chamber never crosses the pressure in mixing chamber, hence there is no inertial effect in question. During the transients the characteristics of primary jet changes throughout owing to the reducing pressure in the mixing chamber, which affects the whole flow field. The experiments agree with the numerical results very well, both, density gradients and secondary chamber pressure are very well matched for the experiments and numerical simulations. The present study clears out the picture of transients occurring in an ejector system till the steady state is achieved.